**Original Manuscript**

***KEGGexpressionMapper*** **allows for analysis of pathways over multiple conditions by integrating transcriptomics and proteomics measurements**

Thomas Nussbaumer[1#], Julia Polzin[2#], and Alexander Platzer[3]

[1] Computational Biology and Systems Biology, Department of Microbiology and Ecosystem Science, Vienna, 1090, Austria.
[2] Division of Microbial Ecology, Department of Microbiology and Ecosystem Science, Vienna, 1090, Austria.
[3] University Clinic for Internal Medicine III, Department of Rheumatology, Medical University Vienna, Waehringer Guertel 18-20, Vienna, Austria

# Joint First Authors

Corresponding author:

Thomas Nussbaumer
thomas.nussbaumer@univie.ac.at

**Abstract**

*Motivation*: In transcriptomic and proteomics-based studies, the abundance of genes is often compared to functional pathways such as the Kyoto Encyclopaedia at Genes and Genomes (KEGG) to identify active metabolic processes. Even though a plethora of tools allow to analyze and to compare '*omics*' data in respect to KEGG pathways, the analysis of multiple conditions is often limited to only a defined set of conditions. Furthermore, for transcriptomic datasets, it is crucial to compare the entire set of pathways in order to obtain a global overview of the species' metabolic functions.

*Results*: Here, we present the tool *KEGGexpressionMapper*, a module, that is implemented in the programming language R. The module allows to highlight the expression of transcriptomic or proteomic measurements in various conditions on pathways and incorporates methods to analyze gene enrichment analyses and expression clustering in time series data. *KEGGexpressionMapper* supports time series data from transcriptomic or proteomics measurements from different individuals. As the tool is implemented in the scripting language R, it can be integrated into existing analysis pipelines to obtain a global overview of the dataset. The R package can be downloaded from https://github.com/nthomasCUBE/KEGGexpressionmapper.

**Introduction**

In transcriptomic- and proteomics-oriented studies, it is often crucial to compare genes to pathway databases such as the Kyoto Encyclopaedia of Genes and Genomes (KEGG) [1], BioCyc [2] and Reactome [3] to obtain valuable information about the functional context of candidate genes and furthermore to determine which pathways are active in the species of interest. This is especially important, when different conditions e.g. different tissues, time series data sets or genotypes are compared. With KEGG being one of the most popular resources for pathway analysis, Pathview [4] and KEGGdb [5] allow to analyze the expression of genes in the context of KEGG pathways. KEGGscape [6] offers also ways to modify or to extend specific pathways, when obvious errors in the pathways should be corrected.

Apart from mapping genes to pathways, it also needs to obtain the functional meaning of these genes. For gene enrichment analysis, GOstats [7] is commonly used to analyze the enrichment of gene ontology terms or certain databases such as DAVID [8] or PIANO [9]. However, these resources are often limited to already well-analysed model species such as human or *Arabidopsis*, whereas non-model organisms can only be compared by applying homology searches against them. However, for species, for which a transcriptome assembly was established and no reference sequence exists and also closely related model species are lacking, only limited options are available.

All of the aforementioned resources for pathway analysis and for gene enrichment analysis are relevant and helpful and provide users with means to explore the context of their candidate genes. However, most tools consider only one condition in order to illustrate the expression or abundance of genes or proteins or focus on only one aspect. Furthermore, tools often lack an option for systematic pathway analyses or require programming skills and knowledge of how to integrate resources by focusing either on the visualization of the KEGG pathways or the enrichments of gene sets. Therefore, we have included publicly available tools such as GOstats to enable analyses of gene sets, that are significantly overrepresented in a particular condition. These gene list can be then extracted and exported into separate files by *KEGGexpressionMapper* and manually inspected and used to enter further empirical analysis.

**Materials and Methods**

*Installation*

*KEGGexpressionMapper* was tested under Linux (Fedora distribution). For Windows, it needs to install the program 'wget' and define it as system variable.

*Extraction of KEGG pathways and integration of expression information*

At first, a user defines the species of interest. Then, all available pathways, containing at least one gene from the species of interest are extracted from KEGG. For each pathway, the gene assignments are taken from the HTML output and customized R functions are used to extract the coordinates of each KEGG compound in a particular pathway. In addition, respective schematic pathway illustrations in the PNG format are downloaded from the KEGG website and the expression values or abundances of a gene or proteins are highlighted by choosing distinct color grades for each quantile by considering the PNG package in R. The expression information or protein abundance for the entire set can be provided via a text file or is taken from an Excel file. A user has to provide a mapping between genes and KEGG orthologous, which can be computed e.g. by considering the KAAST server [10] in a previous step, where proteins of the genome are compared to KEGG orthologous via a bi-directional best hits option, requiring that a complete genome is used as input. The user provides expression measurements for multiple conditions, where values need to be separated by a tabulator. Pathway illustrations in the PNG format are used and modified with help of the PNG package in R which enables the manipulation of PNG illustrations. A user can also include information of candidate genes (e.g. differentially expressed genes) leading to additional output files, which list the overrepresented pathways and functional enrichments for these gene lists. We have also added a color bar for each of the plots to make expression measurements for a pathway comparable to the overall expression of the genes.

*Integration of tools for gene enrichment analysis*

We have integrated the gene enrichment analysis tool GOstats to report enrichments based on an adjusted P value of < 0.05 for each of the three categories, such as Molecular Functions, Biological Process and Chemical Processes. A user can adjust the cut-off for the selection of the adjusted p-value. For each condition, the functional enrichments are then additionally exported into a result Excel. In addition, we highlight genes, that are belong to enriched categories for each of the conditions in the visualisation.

*Over-representation of pathways containing more genes as expected*

We have integrated an additional method to detect pathways, where certain genes are more abundant in a pathway than expected from the total number. Therefore, we make use of a Fisher-test and use multiple testing correcting based on the Benjamini-Hochberg (BH) correction.

**Results**

*Workflow of the KEGGexpressionMapper and comparison to other tools*

Figure 1 depicts the workflow of the *KEGGexpressionMapper*: After the R package is loaded, at first, a user integrates the expression information and optionally, the information of differentially expressed genes as well as the mapping of genes to functional domains. These information is then used to calculate over-representations of domains by considering the tool GOstats. If time-series data is used, we make use of *EBSeqHMM* [11], which allows to group genes into expression profiles. In *EBSeqHMM*, genes are grouped into an expression profile. In order to be assigned to an expression profile, these genes need to show at least one significantly different change in the abundancy between two consecutive time points. If the data represents different individuals, but no time-series data, then the amount of genes of expressed genes is compared for each pathway to the expected number. For genes, that are significantly over-represented in a pathway, these information is integrated into the output of the *KEGGexpressionMapper*. The expression profiles that were provided by *EBSeqHMM* are also used to perform gene ontology term enrichments. In addition, if a gene belongs to an expression profile, these genes are then highlighted on the pathway illustration in a separate figure. *KEGGexpressionMapper* allows to visualize various amounts of conditions and by providing descriptive output files and by a central mapping file to integrate KEGG pathways with expression or protein information.

We illustrate the distinct features of the *KEGGexpressionMapper* in comparison to two other tools, that also provide ways to compare gene lists: DAVID [12] and Piano [9] (Table 1). Whereas DAVID allows performing functional annotations (such as gene-annotations, BioCarta [13] & KEGG mapping), functional classifications, gene ID conversion; *KEGGexpressionMapper* otherwise allows to perform analysis for novel genomes, allows to perform KEGG mapping and provides enrichments analysis. Piano otherwise performs gene set analysis and includes various statistical methods but does not allow comparisons to the KEGG pathways of unknown species. In order to demonstrate the functionality of the tool, we have shown two examples. First, we analysed the symbiotic relations of the clam *Loripes orbiculatus* to an endosymbiont by using five different individuals [14], in the second example, we have used the expression changes in the time-series data of wheat (*Triticum aestivum*) to a fungal pathogen (*Fusarium graminearum*) leading to massive loss of yield by leading to Fusarium head blight (FHB) [15].

*Using KEGGexpressionMapper to study endosymbionts*

The bivalve *Loripes orbiculatus* has an exclusive mutualistic association with a single gammaproteobacterial endosymbiont [14]. For this eukaryote, five different metaproteomics datasets were integrated into the *KEGGexpressionMapper* and allowed to compare the abundancy between the different individuals [14]. Figure 2 illustrates the glycolysis/gluconeogenesis as glucose metabolic pathways, because the endosymbionts converts inorganic substances to feed the host. The colored boxes depict the expression level compared to the global expression and can help to assess whether the direction of the pathway points at breaking down glucose to pyruvate to gain energy in a catabolic reaction or in the reverse direction of forming glucose from non-carbohydrates in an anabolic way. In addition, genes, that are differentially expressed are surrounded by a bounding box, if these represent differentially expressed genes. Furthermore, we provide lists with the functional overrepresentations for each individual in the Excel format by running GOstats, providing in the default mode the gene ontology terms that are below a significance level of 0.1 (see Material and Methods).

*Usage of KEGGexpressionMapper for the study of timeseries data*

We also made use of the data from a time series study, to analyze the interactions between hexaploid bread wheat (*Triticum aestivum*) and Fusarium (*Fusarium graminearum*) under five conditions to demonstrate the feature of *KEGGexpressionMapper* for analysing time series data against KEGG pathways using *EBSeqHMM* [11]. *KEGGexpressionMapper* allowed to assign genes into expression profiles and then compares the amount of genes per profiles between the different conditions. For our example, these are two genotypes, that are highly or less susceptible to Fusarium infection (Figure 3). *KEGGexpressionMapper* thereby allows to assign genes to their expression profiles and provides these gene lists in a separate and cumulative Excel files together with functional enrichments. By using *KEGGexpressionMapper*, we can now easily obtain the genes, that increase in each of the time points and which are also linked to existing KEGG pathways.

**Discussion**

In current transcriptome and proteomics analysis it is important to have access to tools, that allow to systematically analyze the functional repertoire of a species in order to describe novel biology. This is especially different considering genomes, that have no well-analyzed model species. To this end, we have developed *KEGGexpressionMapper*, that allows proteins, that were mapped to KEGG orthologous, to compare the transcriptome and proteomics abundances in pathways. In addition, to find differences between different conditions, such as different species that are more or less effective in particular traits, such as nitrogen-fixation or are more or less resistant to Fusarium infection. With help of *KEGGexpressionMapper*, this allows to generate a set of genes and pathways, that can be then analysed in more detail for more specific gene-based studies.

**Conclusion**

The *KEGGexpressionMapper* allows to visualize a varying amount of conditions in transcriptome and proteome studies to simultaneously compare different treatments, e.g. abiotic or biotic stress to the mock conditions. The output files allow to shortlist pathways and genes, that can be more intensively analysed subsequently. This tool might be helpful for transcriptome studies, where expression is analysed under a certain contrast or for proteomics studies, where it is important, which metabolic pathways are particularly active.

**Acknowledgements**

We acknowledge Professor Dr. Thomas Rattei and Dr. Craig Herbold for valuable comments on the tool and improvements.

**recombination and unique candidate genes in the divergent haplotype**

**encoding Fhb1, a major Fusarium head blight resistance locus in wheat**.

*Theor Appl Genet* 2016, **129**(8):1607-1623.

Figure 1.

**Workflow of the KEGGexpressionMapper**

The KEGGexpressionMapper uses the inputs of expression information, functional domains and integrates time-series or data from various conditions (e.g. various individuals) and runs EBSeq-HMM and GOstats and produces multiple outputs such as transcriptomics and proteomics informed extended pathway visualisations.

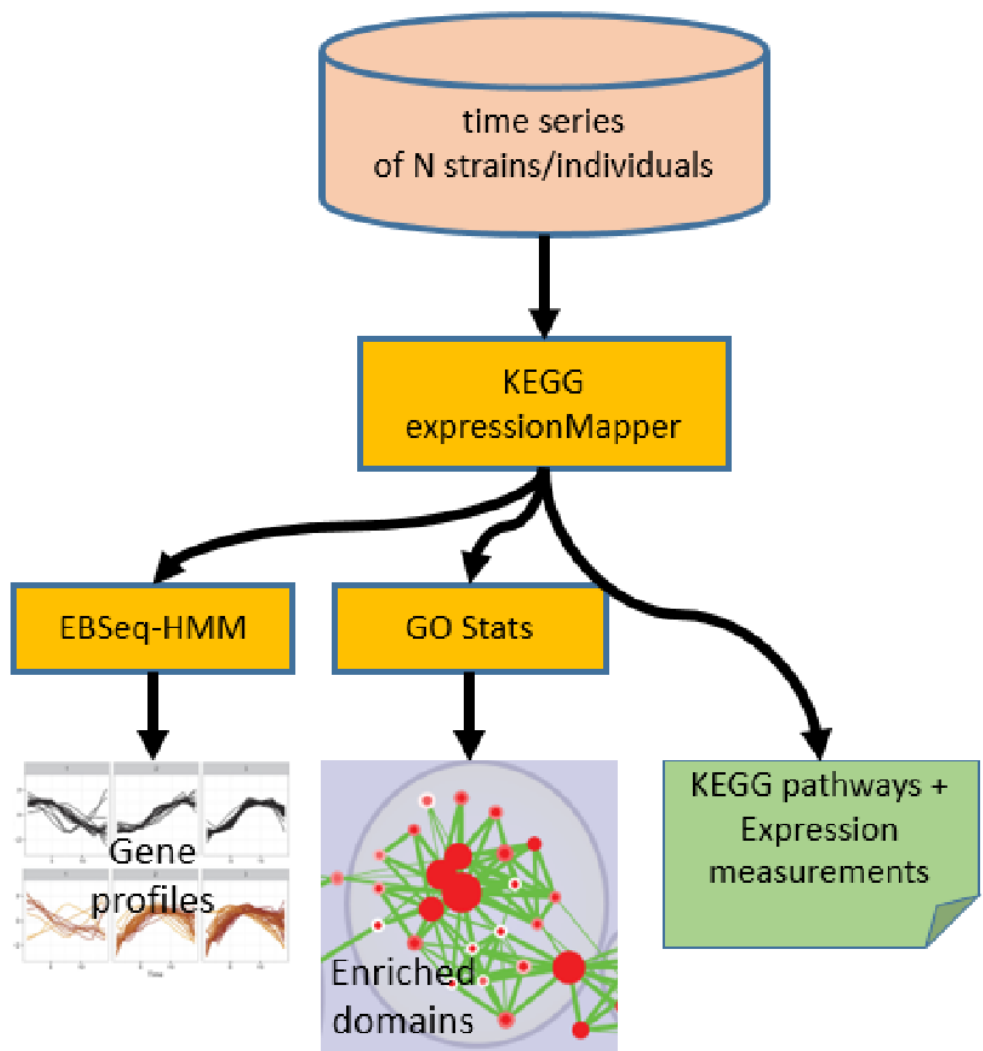

Figure 2.

**Expression-informed module of KEGGexpressionMapper**

Illustration of the host Loripes orbiculatus with expression from the five different individuals after expression information were added. Red colors represent the 25% highest expressed genes, blue the 25% lowest expressed genes.

Figure 3.

## Timeseries data visualisation in KEGGexpressionMapper

Timeseries data from the wheat-Fusarium infection visualized in the KEGGexpressionMapper.

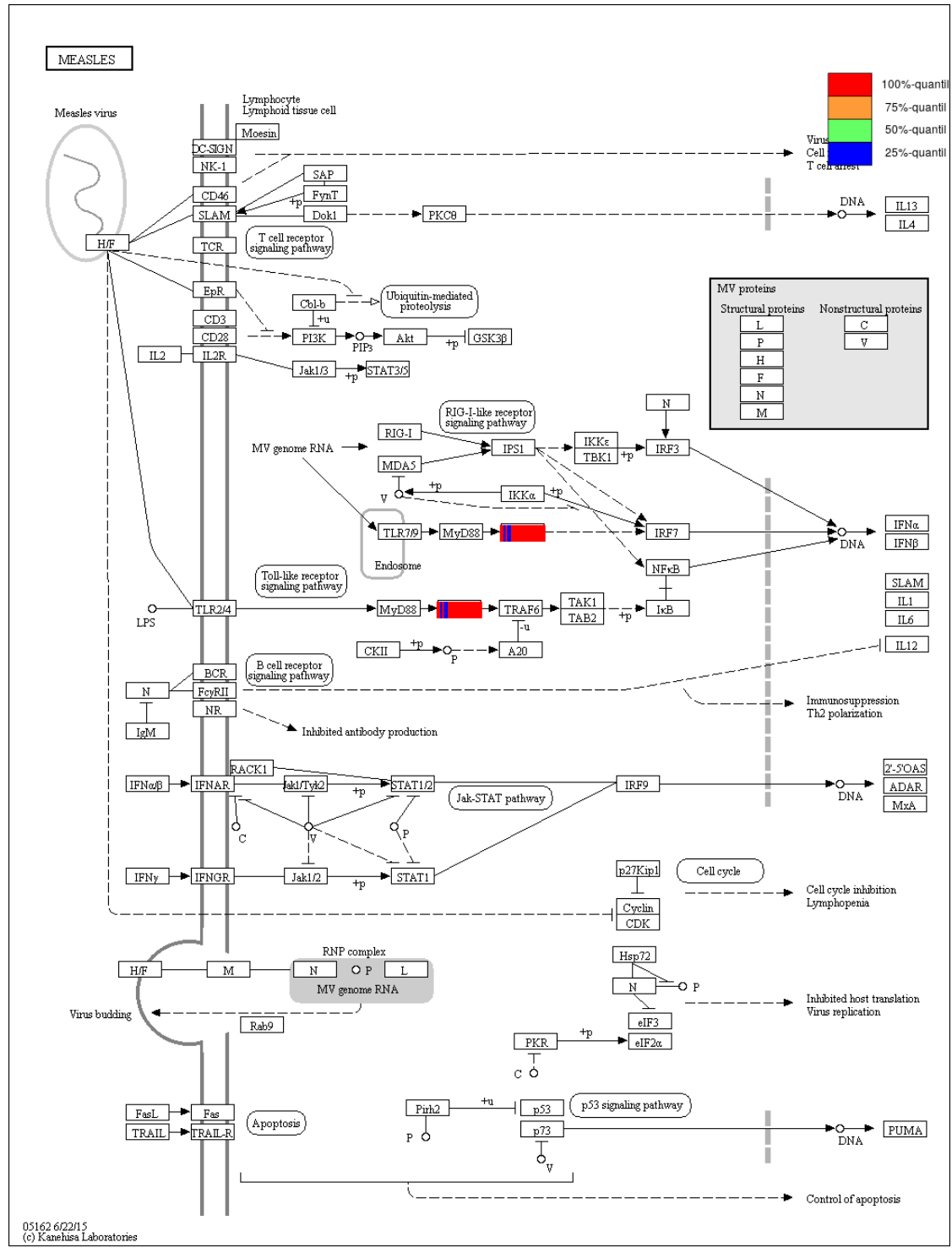